\documentclass[a4paper,12pt]{article}
\usepackage{graphicx}
\usepackage{amsthm}
\usepackage{amsfonts}
\usepackage{amsmath,latexsym}
\usepackage{amssymb}
\usepackage{hyperref}
\usepackage{tabularx}
\newcolumntype{M}{>{$}c<{$}}
\usepackage[matrix,arrow,curve]{xy}

\numberwithin{equation}{section} \numberwithin{figure}{section}
\numberwithin{table}{section}

\def\papertitlepage{\baselineskip 3.5ex\thispagestyle{empty}}
\def\Title#1{\baselineskip 1cm \vspace{1.5cm}%
  \begin{center}{\Large\bf #1}\end{center}\vspace{0.5cm}}
\def\Authors#1{\begin{center}\renewcommand{\thefootnote}{\fnsymbol{footnote}}{\it #1}\end{center}}
\def\Abstract{\vspace{1.0cm}%
  \begin{center}{\large\bf Abstract}\end{center}}

\renewenvironment{thebibliography}{\pagebreak[3]\par\vspace{0.6em}
\begin{flushleft}{\large \bf References}\end{flushleft}
\vspace{-1.0em}

\begin{enumerate}\if@twocolumn\baselineskip=0.6em\itemsep -0.2em
\else\itemsep -0.2em\fi\labelsep 0.1em}{\end{enumerate} }

\begin{document}
{\papertitlepage \vspace*{0cm} {\hfill
\begin{minipage}{4.2cm}
CCNH-UFABC 2017\par\noindent November, 2017
\end{minipage}}
\Title{Comments on real tachyon vacuum solution without square
roots}
\Authors{{\sc E.~Aldo~Arroyo${}$\footnote{\tt
aldo.arroyo@ufabc.edu.br}}
\\
Centro de Ci\^{e}ncias Naturais e Humanas, Universidade Federal do ABC \\[-2ex]
Santo Andr\'{e}, 09210-170 S\~{a}o Paulo, SP, Brazil ${}$ }
} 

\vskip-\baselineskip
{\baselineskip .5cm \Abstract We analyze the consistency of a
recently proposed real tachyon vacuum solution without square
roots in open bosonic string field theory. We show that the
equation of motion contracted with the solution itself is
satisfied. Additionally, by expanding the solution in the basis of
the curly $\mathcal{L}_0$ and the traditional $L_0$ eigenstates,
we evaluate numerically the vacuum energy and obtain a result in
agreement with Sen's conjecture.
 }
\newpage
\setcounter{footnote}{0}
\tableofcontents

\section{Introduction}
In open string field theory \cite{Witten:1985cc}, we say that a
string field $\Psi$ is real if obeys the following reality
condition
\begin{align}
\label{intro1} \Psi^{\ddagger}  = \Psi,
\end{align}
where the double dagger denotes a composition of Hermitian and BPZ
conjugation introduced in Gaberdiel and Zwiebach's seminal work
\cite{Gaberdiel:1997ia}.

Analytic tachyon vacuum solutions that satisfy the above reality
condition (\ref{intro1}) exist in the literature
\cite{Schnabl:2005gv,Schnabl:2010tb}, however they carry some
technical complications. For instance, Schnabl's original solution
is real, but has some subtleties, the solution contains a
singular, projector-like state known as the phantom term
\cite{Erler:2012qr}.

Solutions without the phantom term, known as simple solutions or
Erler-Schnabl's type solutions have been proposed
\cite{Erler:2009uj,Arroyo:2010fq,Zeze:2010sr,Arroyo:2010sy,Erler:2012qn},
but they often fail to satisfy the reality condition. By
performing a gauge transformation over a non-real simple solution,
a real phantom-less solution has been constructed in reference
\cite{Erler:2009uj}. However, as noted in reference
\cite{Jokel:2017vlt}, the cost of having this real solution is the
introduction of somewhat awkward square roots.

It would be desirable to have a solution that is both real and
simple, namely without square roots and phantom terms. This is
precisely the issue that has been studied in a recent paper
\cite{Jokel:2017vlt}, where the author has presented an
alternative prescription to obtain a real solution from a non-real
one which does not make use of a similarity transformation.
Basically, it has been shown that given a tachyon vacuum solution
$\Upsilon$ together with its corresponding homotopy operator $A$
\cite{Ellwood:2001ig,Ellwood:2006ba,Inatomi:2011xr}, the string
field defined by $ \Phi = \text{Re}(\Upsilon) +
\text{Im}(\Upsilon) \, A \, \text{Im}(\Upsilon)$ is a real
solution for the tachyon vacuum.

Applying this prescription for the case of the non-real
Erler-Schnabl's tachyon vacuum solution \cite{Erler:2009uj}
\begin{align}
\label{intro2} \Phi_{\text{Er-Sch}}= c (1+K) Bc\frac{1}{1+K} ,
\end{align}
the corresponding real solution \cite{Jokel:2017vlt} has been
constructed
\begin{align}
\label{intro3} \Phi  = \frac{1}{4}\Big(\frac{1}{1+K}c +
c\frac{1}{1+K} + c \frac{B}{1+K} c +\frac{1}{1+K}c\frac{1}{1+K}
\Big) + Q_B\text{-exact terms},
\end{align}
where the $Q_B\text{-exact terms}$ are given by
\begin{align}
\label{intro4}  \frac{1}{2} \big[ Q_B(Bc)\frac{1}{1+K} +
\frac{1}{1+K}Q_B(Bc) \big] + \frac{1}{4}
\frac{1}{1+K}Q_B(Bc)\frac{1}{1+K}.
\end{align}
For this real solution the corresponding energy has been computed
and shown that the value is in agreement with the value predicted
by Sen's conjecture \cite{Sen:1999mh,Sen:1999xm}.

Nevertheless, for the evaluation of the energy, the equation of
motion contracted with the solution itself was simply assumed to
be satisfied. In this paper, we compute the cubic term of the
action for the real solution (\ref{intro3}) and discuss the
validity of the previous assumption. Additionally, by expanding
the solution in the basis of curly $\mathcal{L}_0$ eigenstates, we
evaluate the energy numerically and obtain a result in agreement
with Sen's conjecture. Since the numerical evaluation of the
energy by means of the curly $\mathcal{L}_0$ level expansion of
the solution is not a trivial task, in order to automate the
computations of relevant correlation functions defined on the
sliver frame, we have developed conservation laws.

This paper is organized as follows. In section 2, we evaluate the
cubic term of the action for the real solution and test the
validity of the equation of motion when contracted with the
solution itself. In section 3, in order to automate the
computations involved in the numerical evaluation of the energy
associated with the solution, we developed conservation laws for
operators defined on the sliver frame. In sections 4, and 5, we
compute the energy by means of the curly $\mathcal{L}_0$ and the
standard Virasoro $L_0$ level expansion of the solution and after
using Pad\'{e} approximants we show that the numerical results
obtained for the energy are in agreement with Sen's conjecture. In
section 6, a summary and further directions of exploration are
given.

\section{Computing the cubic term for the real solution}
In reference \cite{Jokel:2017vlt}, a new real solution for the
tachyon vacuum has been proposed. This solution in the $KBc$
subalgebra \cite{Erler:2006hw,Okawa:2006vm} takes the form
\begin{align}
\label{realplusbrst} \Phi  = \frac{1}{4}\Big(\frac{1}{1+K}c +
c\frac{1}{1+K} + c \frac{B}{1+K} c +\frac{1}{1+K}c\frac{1}{1+K}
\Big) + Q_B\text{-exact terms},
\end{align}
By evaluating the kinetic term of the action, it has been shown
that the energy
\begin{align}
\label{En1} E(\Phi)  = \frac{1}{6} \text{tr}[\Phi Q_B \Phi]
\end{align}
associated with the solution (\ref{realplusbrst}) correctly
reduces to a value which is in accordance with Sen's conjecture.

However, to derive the above equation (\ref{En1}) for the energy,
it has been assumed that the equation of motion holds when
contracted with the solution itself. We know from experience with
other solutions
\cite{Arroyo:2010sy,Okawa:2006vm,Fuchs:2006hw,Arroyo:2009ec} that
this assumption is not a trivial one. In general, a priori there
is no justification for assuming the validity of
\begin{align}
\label{Eqmot1} \text{tr}[ \Phi Q_B\Phi + \Phi \Phi \Phi ] = 0
\end{align}
without an explicit calculation. Therefore the cubic term of the
action must be evaluated.

The computation of the kinetic term has been already done in
reference \cite{Jokel:2017vlt} given as a result
\begin{align}
\label{Kin1} \text{tr}[\Phi Q_B \Phi]= - \frac{3}{\pi^2}.
\end{align}
Thus, for equation (\ref{Eqmot1}) to be valid, we must show that
\begin{align}
\label{Cubin1} \text{tr}[ \Phi \Phi \Phi]= \frac{3}{\pi^2}.
\end{align}

To explicitly compute this cubic term, we need to include the
$Q_B\text{-exact terms}$ of the real solution
(\ref{realplusbrst}). Recall that these terms were not necessary
in the evaluation of the kinetic term. The $Q_B\text{-exact
terms}$ in (\ref{realplusbrst}) are given by
\begin{align}
\label{Brste1}  \frac{1}{2} \big[ Q_B(Bc)\frac{1}{1+K} +
\frac{1}{1+K}Q_B(Bc) \big] + \frac{1}{4}
\frac{1}{1+K}Q_B(Bc)\frac{1}{1+K}.
\end{align}

Inserting the solution (\ref{realplusbrst}) which includes the
$Q_B\text{-exact terms}$ (\ref{Brste1}) into the cubic interaction
term $\text{tr}[ \Phi \Phi \Phi]$, after a lengthy algebraic
manipulations, we arrive to
\begin{align}
\label{Cubin3}\text{tr}[ \Phi \Phi \Phi] =\text{tr}\Big[ -&\frac{1}{16} cKc\frac{1}{1+K}c\frac{1}{(1+K)^2}-\frac{3}{16} cKc\frac{1}{1+K}c\frac{1}{1+K}  \nonumber\\
+&\frac{1}{16} B\frac{1}{(1+K)^2}cKc\frac{1}{(1+K)^2}cKc+\frac{3}{16} B\frac{1}{(1+K)^2}cKc\frac{1}{1+K}cKc  \nonumber\\
+&\frac{1}{8} B\frac{1}{1+K}cKc\frac{1}{(1+K)^2}cKc+\frac{1}{8} B\frac{1}{1+K}cKc\frac{1}{1+K}cKc  \nonumber\\
-&\frac{1}{16}
B\frac{1}{1+K}c\frac{1}{1+K}cKc\frac{1}{1+K}c+\frac{1}{16}
B\frac{1}{1+K}c\frac{1}{1+K}c\frac{1}{1+K}cKc \Big].
\end{align}

All the correlators appearing in the evaluation of the cubic term
(\ref{Cubin3}) can be computed by means of the following basic
correlators
\begin{align}
\label{corref13ccc} \text{tr} \big[  c e^{-t_1 K} c e^{-t_2 K} c
e^{-t_3 K} \big] = \frac{\left(t_1+t_2+t_3\right){}^3 \sin
\left(\frac{\pi  t_2}{t_1+t_2+t_3}\right) \sin \left(\frac{\pi
t_3}{t_1+t_2+t_3}\right) \sin \left(\frac{\pi
    \left(t_2+t_3\right)}{t_1+t_2+t_3}\right)}{\pi ^3} , \\
\text{tr} \big[ B e^{-t_1 K} c e^{-t_2 K} c e^{-t_3 K} c e^{-t_4
K} c  \big] =\frac{s^2 (t_2+t_3+t_4)}{\pi ^3} \sin (\frac{\pi
t_3}{s}) \sin (\frac{\pi t_4}{s}) \sin
  (\frac{\pi  (t_3+t_4)}{s}) \nonumber \\
- \frac{s^2 (t_3+t_4)}{\pi ^3}\sin (\frac{\pi t_4}{s}) \sin
(\frac{\pi
   (t_2+t_3)}{s}) \sin (\frac{\pi
   (t_2+t_3+t_4)}{s}) \nonumber \\
\label{corref13bcccc}    + \frac{s^2 t_4}{\pi ^3} \sin (\frac{\pi
t_2}{s}) \sin (\frac{\pi
   (t_3+t_4)}{s}) \sin (\frac{\pi
   (t_2+t_3+t_4)}{s}),
\end{align}
where $s=t_1+t_2+t_3+t_4$.

For instance, employing the correlator (\ref{corref13ccc}), let us
explicitly compute the correlator
\begin{align}
\text{tr}\Big[
    cKc\frac{1}{1+K}c\frac{1}{(1+K)^2}  \Big] &= -
    \int_{0}^{\infty}dt_1dt_2 \, t_2 e^{-t_1-t_2} \partial_s \text{tr}\Big[
    ce^{-s K}ce^{-t_1 K}ce^{-t_2 K}  \Big]\Big{|}_{s=0} \nonumber
    \\
\label{integralccc1} &=  -
    \int_{0}^{\infty}dt_1dt_2 \, t_2 e^{-t_1-t_2}
    \frac{\left(t_1+t_2\right){}^2 \sin \left(\frac{\pi  t_1}{t_1+t_2}\right)
    \sin \left(\frac{\pi  t_2}{t_1+t_2}\right)}{\pi ^2} .
\end{align}
To evaluate the above double integral, we perform the change of
variables $t_1 \rightarrow  u v$, $t_2 \rightarrow  u - u v$,
$\int_{0}^{\infty} dt_1 dt_2 \rightarrow \int_{0}^{\infty} du
\int_{0}^{1} dv \, u  $, so that from equation
(\ref{integralccc1}), we obtain
\begin{align}
\text{tr}\Big[
    cKc\frac{1}{1+K}c\frac{1}{(1+K)^2}  \Big] &= \int_{0}^{\infty} du \int_{0}^{1} dv \, \frac{e^{-u} u^4 (v-1) \sin ^2(\pi  v)}{\pi ^2} \nonumber
    \\
\label{integralccc2} &= -\frac{6}{\pi ^2}.
\end{align}

To compute correlators containing the $B$ string field, we proceed
in the same manner. As an illustration, let us explicitly evaluate
the correlator $\text{tr}\Big[
B\frac{1}{(1+K)^2}cKc\frac{1}{(1+K)^2}cKc
      \Big]$. The integral representation of this correlator is
      given by
\begin{align}
\label{integralbcccc1}    \int_{0}^{\infty}dt_1dt_2 \, t_1 t_2
e^{-t_1-t_2} \partial_{s_1,s_2} \text{tr}\Big[
    B e^{-t_1 K} c e^{-s_1 K} c e^{-t_2 K} c e^{-s_2
K} c   \Big]\Big{|}_{s_1=s_2=0} .
\end{align}
Using the correlator (\ref{corref13bcccc}), from equation
(\ref{integralbcccc1}) we obtain
\begin{align}
\label{integralbcccc2}    \int_{0}^{\infty}dt_1dt_2 \, t_1 t_2
e^{-t_1-t_2} \frac{2 \sin \left(\frac{\pi  t_2}{t_1+t_2}\right)
\left(\left(t_1+t_2\right) \sin \left(\frac{\pi
t_2}{t_1+t_2}\right)-\pi  t_2 \cos
   \left(\frac{\pi  t_2}{t_1+t_2}\right)\right)}{\pi ^2} .
\end{align}
Performing the change of variables $t_1 \rightarrow  u v$, $t_2
\rightarrow  u - u v$, $\int_{0}^{\infty} dt_1 dt_2 \rightarrow
\int_{0}^{\infty} du \int_{0}^{1} dv \, u  $ into the above double
integral (\ref{integralbcccc2}), we get
\begin{align}
\label{integralbcccc3} \int_{0}^{\infty} du \int_{0}^{1} dv
\frac{2 e^{-u} u^4 (1-v) v \sin (\pi  v) (\sin (\pi  v)-\pi  (v-1)
\cos (\pi  v))}{\pi ^2} = \frac{30}{\pi ^4}+\frac{4}{\pi ^2}.
\end{align}
Therefore, we have just shown that
\begin{align}
\label{integralbcccc4}\text{tr}\Big[
B\frac{1}{(1+K)^2}cKc\frac{1}{(1+K)^2}cKc
      \Big] = \frac{30}{\pi ^4}+\frac{4}{\pi ^2}.
\end{align}

In this way, we can calculate all the relevant correlators
appearing in the right hand side of equation (\ref{Cubin3}). Let
us list the results
\begin{align}
\label{Cubin3ter1} \text{tr}\Big[
    cKc\frac{1}{1+K}c\frac{1}{(1+K)^2}  \Big]&= -\frac{6}{\pi ^2}, \\
 \text{tr}\Big[ cKc\frac{1}{1+K}c\frac{1}{1+K}
      \Big]&= -\frac{3}{\pi ^2}, \\
\text{tr}\Big[ B\frac{1}{(1+K)^2}cKc\frac{1}{(1+K)^2}cKc
      \Big] &= \frac{30}{\pi ^4}+\frac{4}{\pi ^2}, \\
\text{tr}\Big[B\frac{1}{(1+K)^2}cKc\frac{1}{1+K}cKc
      \Big] &= \frac{3}{\pi ^2}, \\
\text{tr}\Big[B\frac{1}{1+K}cKc\frac{1}{(1+K)^2}cKc
      \Big] &= \frac{6}{\pi ^2}, \\
\text{tr}\Big[ B\frac{1}{1+K}cKc\frac{1}{1+K}cKc
      \Big] &= \frac{3}{\pi ^2}, \\
      \text{tr}\Big[ B\frac{1}{1+K}c\frac{1}{1+K}cKc\frac{1}{1+K}c
      \Big] &= \frac{15}{\pi ^4}-\frac{4}{\pi ^2}, \\
\label{Cubin3ter2} \text{tr}\Big[
B\frac{1}{1+K}c\frac{1}{1+K}c\frac{1}{1+K}cKc
      \Big] &= -\frac{15}{\pi ^4}-\frac{2}{\pi ^2}.
\end{align}

Employing these results (\ref{Cubin3ter1})-(\ref{Cubin3ter2}) into
equation (\ref{Cubin3}) and adding up all terms, we obtain the
value for the cubic term
\begin{align}
\label{Cubin4} \text{tr}[ \Phi \Phi \Phi] = \frac{3}{\pi^2}.
\end{align}

Since we have explicitly shown that the equation of motion is
satisfied when contracted with the solution itself, i.e. $
\text{tr}[ \Phi Q_B \Phi ]+\text{tr}[ \Phi \Phi \Phi]=0$, it is
guaranteed that the energy associated with the solution
(\ref{realplusbrst}) is directly proportional to the kinetic term
\begin{align}
\label{EnerIntro1} E(\Phi) = -\mathcal{S}[\Phi]
=\frac{1}{2}\text{tr}[ \Phi Q_B \Phi ] + \frac{1}{3}\text{tr}[
\Phi \Phi \Phi] = \frac{1}{6} \text{tr}[ \Phi Q_B \Phi ].
\end{align}

As a second test of consistency, we would like to analyze the
solution from a numerical point of view, in particular, we will be
interested in the numerical evaluation of the kinetic term by
means of the curly $\mathcal{L}_0$ level expansion of the
solution.

As we are going to show, when we insert the curly $\mathcal{L}_0$
level expansion of the solution into the kinetic term, we are
required to evaluate two point vertices for string fields
containing the operators $\hat{\mathcal{L}}$, $\hat{\mathcal{B}}$
and $\tilde c_p$. These two point vertices can be evaluated by
means of the so-called conservation laws which will be studied in
the next section.

\section{Conservation laws and the two point vertex in the sliver frame}
The operators employed in the basis of curly $\mathcal{L}_0$
eigenstates are given in terms of the basic operators
$\hat{\mathcal{L}}$, $\hat{\mathcal{B}}$ and $\tilde c_p$. These
operators are related to the worldsheet energy momentum tensor
$T(z)$, the $b(z)$ and $c(z)$ ghosts fields respectively. We are
going to derive the conservation law for the $\hat{\mathcal{L}}$
operator
\begin{eqnarray}
\label{lhatreal} \hat{\mathcal{L}}= \oint \frac{d z}{2 \pi i}
(1+z^{2}) (\arctan z+\text{arccot} z) \, T(z) \, .
\end{eqnarray}

Using the conformal map $\tilde z= \frac{2}{\pi}\arctan z $, we
can write the expression of the $\hat{\mathcal{L}}$ operator in
the sliver frame
\begin{eqnarray}
\label{lhat2} \hat{\mathcal{L}}= \oint \frac{d \tilde z}{2 \pi i}
\varepsilon (\text{Re} \tilde z) \, \tilde T( \tilde z) \, ,
\end{eqnarray}
where $\varepsilon (x)$ is the step function equal to $\pm 1$ for
positive or negative values of its argument respectively.

For vertex operators $\phi_i$ defined on the sliver frame, the two
functions $f_1$ and $f_2$ which appear in the definition of the
two point vertex $\big\langle f_1 \circ \phi_1(0) f_2 \circ
\phi_2(0)
 \big\rangle$ are given by
\begin{eqnarray}
\label{ff1} f_1(\tilde z_1) &=& \tan \big(\frac{\pi}{2}(1+\tilde z_1)\big) \, , \\
\label{ff2} f_2(\tilde z_2) &=& \tan \big( \frac{\pi}{2}\tilde
z_2\big) \, .
\end{eqnarray}

We need conservation laws such that the operator
$\hat{\mathcal{L}}$ acting on the two point vertex, which we
denote as $\big\langle V_2 \big|$, can be expressed in terms of
non-negative Virasoro modes defined on the sliver
frame\footnote{We are going to use the following notation
$\mathcal{O}^{(i)}$ to refer an operator $\mathcal{O}$ defined
around the $i$-th puncture.}
\begin{eqnarray}
\label{V2Lhat1} \big\langle V_2 \big| \hat{\mathcal{L}}^{(2)} =
\big\langle V_2 \big| \Big(  \sum_{n\geq0}  a_n
\mathcal{L}_n^{(1)} + \sum_{n\geq0 }  b_n \mathcal{L}_n^{(2)}
  \Big) \, ,
\end{eqnarray}
where $a_n$ and $b_n$ are coefficients that will be determined
below.

To derive a conservation law of the form (\ref{V2Lhat1}), we need
a vector field which behaves as $ v^{(2)}(\tilde z_2) \sim
 \varepsilon (\text{Re} \tilde z_2)+O(\tilde z_2)$ around puncture
2, and has the following behavior in the other puncture,
$v^{(1)}(\tilde z_1) \sim O(\tilde z_1)$. A vector field which
does this job is given by
\begin{eqnarray}
\label{vect2} v(z) = (1+z^2) \text{arccot}z .
\end{eqnarray}

The expression of the conservation law for Virasoro modes defined
on the sliver frame is given by\footnote{This formula can be
derived using the general prescription for conservation laws shown
in references \cite{Rastelli:2000iu,Arroyo:2011zt}.}
\begin{eqnarray}
\label{conservaV2} \big\langle V_2 \big| \sum_{j=1}^{2}
\oint_{\mathcal{C}_j} \frac{1}{2 \pi i} v^{(j)}(\tilde z_j) \tilde
T(\tilde z_j) d\tilde z_j = 0 \, ,
\end{eqnarray}
where $v^{(j)}(\tilde z_j) = ( \partial_{\tilde z_j}f_{j} (\tilde
z_j))^{-1} v(f_j (\tilde z_j)) $, and $\mathcal{C}_j$ is a closed
contour which encircles the $j$-puncture.

Using equations (\ref{ff1}), (\ref{ff2}) and (\ref{vect2}) into
the definition $v^{(j)}(\tilde z_j) = (
\partial_{\tilde{z}_j}f_{j} (\tilde{z}_j))^{-1} v(f_j (\tilde z_j))$
of the vector fields $v^{(1)}(\tilde z_1)$ and $v^{(2)}(\tilde
z_2)$, we find that
\begin{eqnarray}
v^{(1)}(\tilde z_1) &=&-\tilde z_1 \\
v^{(2)}(\tilde z_2) &=&  \varepsilon (\text{Re} \tilde z_2)-\tilde
z_2 .
\end{eqnarray}
Due to the presence of the step function we see that the vector
field $v^{(2)}(\tilde z_2)$ is discontinuous around puncture 2,
since we are interested in the conservation law of the operator
defined in equation (\ref{lhat2}), this kind of discontinuity is
what we want. Using (\ref{conservaV2}) and noting that integration
amounts to the replacement $v^{(i)}_n \tilde z_i^n \rightarrow
v^{(i)}_n \mathcal{L}^{(i)}_ {n-1}$, we can immediately write the
conservation law
\begin{align}
\label{conservaeq2} \big\langle V_2 \big| \Big(
-\mathcal{L}_0^{(1)} + \hat{\mathcal{L}}^{(2)} -
\mathcal{L}_0^{(2)} \Big) = 0 \, .
\end{align}
We can write this conservation law (\ref{conservaeq2}) in the
standard form as given in equation (\ref{V2Lhat1})
\begin{align}
\label{conservaeq3}\big\langle V_2 \big| \hat{\mathcal{L}}^{(2)} =
\big\langle V_2 \big| \Big( \mathcal{L}_0^{(1)} +
\mathcal{L}_0^{(2)} \Big) \, .
\end{align}
By the symmetry property of the two vertex, the same identity
(\ref{conservaeq3}) holds after replacing (1) $\rightarrow$ (2)
\begin{align}
\label{conservaeq4}\big\langle V_2 \big| \hat{\mathcal{L}}^{(1)} =
\big\langle V_2 \big| \Big( \mathcal{L}_0^{(2)} +
\mathcal{L}_0^{(1)} \Big) \, .
\end{align}

Regarding the conservation law for the $\hat{\mathcal{B}}$
operator, since the $b$ ghost is a conformal field of dimension
two, the conservation laws for operators involving this field are
identical to those for the Virasoro operators
\begin{align}
\label{conservaeq5} \big\langle V_2 \big| \hat{\mathcal{B}}^{(2)}
&= \big\langle V_2 \big| \Big( \mathcal{B}_0^{(1)} +
\mathcal{B}_0^{(2)} \Big) \, , \\
\label{conservaeq6} \big\langle V_2 \big| \hat{\mathcal{B}}^{(1)}
& = \big\langle V_2 \big| \Big( \mathcal{B}_0^{(2)} +
\mathcal{B}_0^{(1)} \Big) \, .
\end{align}

Employing these conservation laws for the operators
$\hat{\mathcal{L}}$ and $\hat{\mathcal{B}}$, together with the
commutator and anti-commutator relations
\begin{align}
\label{com1}
[\mathcal{L}_0^{(i)},\hat{\mathcal{L}}^{(j)}]&=\delta^{ij}\hat{\mathcal{L}}^{(j)},
\;\;\;\;\;
[\mathcal{L}_0^{(i)},\hat{\mathcal{B}}^{(j)}]=\delta^{ij}\hat{\mathcal{B}}^{(j)},
\;\;\;\;\; [\mathcal{L}_0^{(i)},\tilde c_p^{(j)}] =
-\delta^{ij}\,p\, c_p^{(j)}\, , \\
\label{com2}
[\mathcal{B}_0^{(i)},\hat{\mathcal{L}}^{(j)}]&=\delta^{ij}\hat{\mathcal{B}}^{(j)},
\;\;\;\, \{\mathcal{B}_0^{(i)},\hat{\mathcal{B}}^{(j)}\}=0,
\;\;\;\;\;\;\;\;\;\;\; \{\mathcal{B}_0^{(i)},\tilde c_p^{(j)}\} =
\delta^{ij}\, \delta_{0,p} \, ,
\end{align}
we can show that all two point correlation functions involving
string fields constructed out of the operators
$\hat{\mathcal{L}}$, $\hat{\mathcal{B}}$ and $\tilde c_p$ can be
reduced to the evaluation of the following basic correlators
\begin{align}
\label{corccc1} \big\langle V_2 \big|\tilde c_{p_1}^{(2)} \tilde
c_{p_2}^{(2)} \tilde c_{p_3}^{(2)} \big\rangle = \big\langle V_2
\big|\tilde c_{p_1}^{(1)} \tilde c_{p_2}^{(1)} \tilde
c_{p_3}^{(1)} \big\rangle &= \oint \frac{dx_1 dx_2 dx_3}{(2 \pi
i)^3} x^{p_1-2}_1 x^{p_2-2}_2 x^{p_3-2}_3 \big\langle
c(x_1)c(x_2)c(x_3)\big\rangle_{\mathcal{C}_2}, \\
\label{corccc2} \big\langle V_2 \big|\tilde c_{p_1}^{(1)} \tilde
c_{p_2}^{(2)} \tilde c_{p_3}^{(2)} \big\rangle = \big\langle V_2
\big|\tilde c_{p_2}^{(1)} \tilde c_{p_3}^{(1)} \tilde
c_{p_1}^{(2)} \big\rangle &= \oint \frac{dx_1 dx_2 dx_3}{(2 \pi
i)^3} x^{p_1-2}_1 x^{p_2-2}_2 x^{p_3-2}_3 \big\langle
c(x_1+1)c(x_2)c(x_3)\big\rangle_{\mathcal{C}_2},
\end{align}
where the correlator $\big\langle
c(x)c(y)c(z)\big\rangle_{\mathcal{C}_L}$ in general is given by
\begin{eqnarray}
\label{corccc3}\big\langle c(x)c(y)c(z)\big\rangle_{C_L} =
\frac{L^3}{\pi ^3} \sin \left(\frac{\pi
\left(x-y\right)}{L}\right) \sin
   \left(\frac{\pi  \left(x-z\right)}{L}\right) \sin \left(\frac{\pi
   \left(y-z\right)}{L}\right) .
\end{eqnarray}

To evaluate explicitly the above correlators (\ref{corccc1}) and
(\ref{corccc2}), the following formulas will be very useful
\begin{align}
\label{Fs} S_{a,b}\equiv \oint \frac{dz}{2 \pi i} z^a \sin (b z) =
-\frac{b^{-a-1} \cos \left(\frac{\pi
a}{2}\right)}{\Gamma (-a)}, \\
\label{Fc} C_{a,b}\equiv \oint \frac{dz}{2 \pi i} z^a \cos (b z) =
-\frac{b^{-a-1} \sin \left(\frac{\pi a}{2}\right)}{\Gamma (-a)}.
\end{align}
For instance, let us compute correlator (\ref{corccc1}). Using
(\ref{corccc3}) into equation (\ref{corccc1}), we have
\begin{align}
\big\langle V_2 \big|\tilde c_{p_1}^{(2)} \tilde c_{p_2}^{(2)}
\tilde c_{p_3}^{(2)} \big\rangle = \big\langle V_2 \big|\tilde
c_{p_1}^{(1)} \tilde c_{p_2}^{(1)} \tilde c_{p_3}^{(1)}
\big\rangle =\oint \frac{dx_1 dx_2 dx_3}{(2 \pi i)^3} x^{p_1-2}_1
x^{p_2-2}_2 x^{p_3-2}_3 \big\langle
c(x_1)c(x_2)c(x_3)\big\rangle_{\mathcal{C}_2} = \nonumber \\
= \frac{2}{\pi^3} \oint \frac{dx_1 dx_2 dx_3}{(2 \pi i)^3}
x^{p_1-2}_1 x^{p_2-2}_2 x^{p_3-2}_3 \Big[ \sin \left(\pi
x_1\right) \cos \left(\pi  x_2\right)-\sin \left(\pi  x_1\right)
\cos \left(\pi  x_3\right)  \;\;\; \nonumber  \\ +\sin \left(\pi
x_2\right) \cos \left(\pi  x_3\right)-\sin \left(\pi  x_2\right)
\cos \left(\pi x_1\right) \;\;\;\, \nonumber  \\ \label{Exp1} +
\sin \left(\pi x_3\right) \cos \left(\pi x_1\right)-\sin \left(\pi
x_3\right) \cos \left(\pi x_2\right) \Big].
\end{align}
It is clear that the above equation (\ref{Exp1}) can be written in
terms of the functions (\ref{Fs}) and (\ref{Fc}), so that we
arrive to an explicit expression for the correlator
(\ref{corccc1})
\begin{align}
\big\langle V_2 \big|\tilde c_{p_1}^{(2)} \tilde c_{p_2}^{(2)}
\tilde c_{p_3}^{(2)} \big\rangle = \big\langle V_2 \big|\tilde
c_{p_1}^{(1)} \tilde c_{p_2}^{(1)} \tilde c_{p_3}^{(1)}
\big\rangle = \frac{2}{\pi^3} \Big[ \delta_{1,p_3} S_{p_1-2,\pi}
C_{p_2-2,\pi } -\delta_{1,p_2}C_{p_3-2,\pi } S_{p_1-2,\pi }
\nonumber
\\ \label{Exp2} -\delta_{1,p_3}C_{p_1-2,\pi } S_{p_2-2,\pi }  + \delta_{1,p_1} C_{p_3-2,\pi } S_{p_2-2,\pi }+\delta_{1,p_2}C_{p_1-2,\pi } S_{p_3-2,\pi
   }-\delta_{1,p_1}C_{p_2-2,\pi } S_{p_3-2,\pi }\Big].
\end{align}
In the same way, we can also derive the explicit expression for
the correlator (\ref{corccc2})
\begin{align}
\big\langle V_2 \big|\tilde c_{p_1}^{(1)} \tilde c_{p_2}^{(2)}
\tilde c_{p_3}^{(2)} \big\rangle = \big\langle V_2 \big|\tilde
c_{p_2}^{(1)} \tilde c_{p_3}^{(1)} \tilde c_{p_1}^{(2)}
\big\rangle = \frac{2}{\pi^3} \Big[ \delta_{1,p_2} S_{p_1-2,\pi}
C_{p_3-2,\pi } -\delta_{1,p_3}C_{p_2-2,\pi } S_{p_1-2,\pi }
\nonumber
\\ \label{Exp3} -\delta_{1,p_2}C_{p_1-2,\pi } S_{p_3-2,\pi }  + \delta_{1,p_3} C_{p_1-2,\pi } S_{p_2-2,\pi }+\delta_{1,p_1}C_{p_3-2,\pi } S_{p_2-2,\pi
   }-\delta_{1,p_1}C_{p_2-2,\pi } S_{p_3-2,\pi }\Big].
\end{align}

To evaluate the kinetic term $\text{tr}[\Phi \, Q_B \Phi]$ for a
string field $\Phi$ expanded in the basis of curly $\mathcal{L}_0$
eigenstates, it will be convenient to write the kinetic term in
the language of a two point vertex
\begin{align}
\text{tr}[\Phi \, Q_B  \Phi] = \big\langle V_2 \big| \Phi^{(1)}Q_B
\Phi^{(2)} \big \rangle.
\end{align}
Note that in addition to the conservation laws, we will be
required to know the action of the BRST charge $Q_B$ on the
operators $\hat{\mathcal{L}}$, $\hat{\mathcal{B}}$ and $\tilde
c_p$
\begin{align}
\label{BRSTact1} [Q_B, \hat{\mathcal{L}}^{(j)}] = 0, \;\;\;\;\;
\{Q_B, \hat{\mathcal{B}}^{(j)}\} =\hat{\mathcal{L}}^{(j)},
\;\;\;\;\; \{Q_B, \tilde c_p^{(j)}\} = \sum_{k=-\infty}^{\infty}
(1-k) \tilde c_{p-k}^{(j)} \tilde c_k^{(j)}.
\end{align}

As an illustration of the use of conservation laws, we are going
to compute a particular correlator involving the operators
$\hat{\mathcal{B}}$ and $\hat{\mathcal{L}}$. We choose, as an
example, the following string fields
\begin{align}
\label{ph1} \phi = \hat{\mathcal{B}} \hat{\mathcal{L}} \tilde c_0
\tilde c_1|0\rangle, \;\;\;\;\;\; \psi = \tilde c_1|0\rangle.
\end{align}
Using these string fields, let us evaluate the correlator
\begin{align}
\label{corexe1} \text{tr}[\phi Q_B \psi] =\big\langle V_2 \big|
\phi^{(1)}Q_B \psi^{(2)} \big \rangle.
\end{align}
Inserting equation (\ref{ph1}) into equation (\ref{corexe1}) and
using (\ref{BRSTact1}), we obtain
\begin{align}
\label{corexe2} \text{tr}[\phi Q_B \psi] =-\big\langle V_2 \big|
\hat{\mathcal{B}}^{(1)} \hat{\mathcal{L}}^{(1)} \tilde c_0^{(1)}
\tilde c_1^{(1)}
 \tilde c_0^{(2)} \tilde c_1^{(2)} \big \rangle.
\end{align}
Using the conservation law (\ref{conservaeq6}) and the
anti-commutator relations (\ref{com2}), from equation
(\ref{corexe2}) we get
\begin{align}
\label{corexe3} \text{tr}[\phi Q_B \psi] =-\big\langle V_2 \big|
\hat{\mathcal{L}}^{(1)} \tilde c_0^{(1)} \tilde c_1^{(1)}
 \tilde c_1^{(2)}  \big \rangle -\big\langle V_2 \big|
\hat{\mathcal{L}}^{(1)} \tilde c_1^{(1)} \tilde c_0^{(2)}
 \tilde c_1^{(2)}  \big \rangle -\big\langle V_2 \big|
\hat{\mathcal{B}}^{(1)} \tilde c_0^{(1)} \tilde c_1^{(1)}
 \tilde c_0^{(2)}  \tilde c_1^{(2)} \big \rangle .
\end{align}
Employing the conservation laws (\ref{conservaeq4}),
(\ref{conservaeq6}) and the commutator and anti-commutator
relations (\ref{com1}), (\ref{com2}), from equation
(\ref{corexe3}) we arrive to
\begin{align}
\label{corexe4} \text{tr}[\phi Q_B \psi] = \big\langle V_2 \big|
\tilde c_0^{(1)} \tilde c_1^{(1)}
 \tilde c_1^{(2)}  \big \rangle + \big\langle V_2 \big|
\tilde c_1^{(1)} \tilde c_0^{(2)}
 \tilde c_1^{(2)}  \big \rangle = 2 \big\langle V_2 \big|
\tilde c_1^{(1)} \tilde c_0^{(2)}
 \tilde c_1^{(2)}  \big \rangle = 2 \Big(\frac{4}{\pi ^2}\Big) = \frac{8}{\pi ^2}
 ,
\end{align}
where we have used equation (\ref{Exp3}). These kind of
computations can be automated in a computer. Next, we are going to
apply the results shown in this section to evaluate the kinetic
term by means of the curly $\mathcal{L}_0$ level expansion of the
real solution (\ref{realplusbrst}).

\section{Curly $\mathcal{L}_0$ level expansion analysis of the real solution}

Since the kinetic term does not depend on the $Q_B\text{-exact
terms}$, we are going to consider only the first term of $\Phi$
given in equation (\ref{realplusbrst}). Let us define this term as
\begin{align}
\label{realplusbrstlo} \hat \Phi  = \frac{1}{4}\Big(\frac{1}{1+K}c
+ c\frac{1}{1+K} + c \frac{B}{1+K} c +\frac{1}{1+K}c\frac{1}{1+K}
\Big).
\end{align}
Using the integral representation of $1/(1+K)$
\begin{align}
\label{inK1} \frac{1}{1+K} = \int_0^{\infty} dt \, e^{-t(1+K)} =
\int_0^{\infty} dt \, e^{-t} \Omega^t,
\end{align}
we can write (\ref{realplusbrstlo}) as
\begin{align}
\label{realplusbrstl02} \hat \Phi  =
\frac{1}{4}\Big[\int_0^{\infty} dt \, e^{-t} \Big(\Omega^t c +
c\Omega^t + c \Omega^t B c  \Big) + \int_0^{\infty} ds dt \,
e^{-s-t} \Omega^s c \Omega^t \Big].
\end{align}

By writing the basic string fields $K$, $B$ in terms of the
operators $\hat{\mathcal{L}}$, $\hat{\mathcal{B}}$, and using the
modes $\tilde c_p$ of the ghost field $c(z)$ defined in the
$\tilde z$-conformal frame $\tilde z = \frac{2}{\pi} \arctan z$,
we can show that
\begin{align}
\label{realplusbrstl03} \Omega^{t_1} c \Omega^{t_2} B c
\Omega^{t_3} = & \sum_{n=0}^{\infty} \sum_{p=-\infty}^{1}
\frac{\beta^n}{2n!} (x^{1-p}+y^{1-p}) \hat{\mathcal{L}}^n \tilde
c_p |0\rangle \nonumber \\ &+ \sum_{n=0}^{\infty}
\sum_{p=-\infty}^{1} \sum_{q=-\infty}^{1} \frac{\beta^n}{4 n!}
(x^{1-p}y^{1-q}-x^{1-q}y^{1-p}) \hat{\mathcal{B}}
\hat{\mathcal{L}}^n \tilde c_p \tilde c_q |0\rangle,
\end{align}
where
\begin{eqnarray}
\label{realplusbrstl04} \beta=\frac{1}{2}-\frac{1}{2}(t_1+t_2+t_3)
, \;\;\; x=\frac{1}{2}(t_3-t_1-t_2), \;\;\;
y=\frac{1}{2}(t_2+t_3-t_1).
\end{eqnarray}

Employing equation (\ref{realplusbrstl03}), it is possible to
derive the curly $\mathcal{L}_0$ level expansion of the string
field defined in equation (\ref{realplusbrstl02}). As a
pedagogical illustration, let us explicitly compute the curly
$\mathcal{L}_0$ level expansion of the last term appearing on the
right hand side of equation (\ref{realplusbrstl02})
\begin{align}
\label{realplusbrstl05} \int_0^{\infty} ds dt \, e^{-s-t} \Omega^s
c \Omega^t = \sum_{n=0}^{\infty} \sum_{p=-\infty}^{1}
\int_0^{\infty} ds dt \, e^{-s-t} \frac{\beta^n}{2n!}
(x^{1-p}+y^{1-p}) \hat{\mathcal{L}}^n \tilde c_p |0\rangle ,
\end{align}
where in this case
\begin{eqnarray}
\label{realplusbrstl06} \beta=\frac{1}{2}-\frac{1}{2}(s+t) ,
\;\;\;\;\;\; x=y=\frac{1}{2}(t-s).
\end{eqnarray}

As we can see from equations (\ref{realplusbrstl05}) and
(\ref{realplusbrstl06}), we are required to evaluate the following
double integral
\begin{align}
\label{realplusbrstl07} \int_0^{\infty} ds dt \, e^{-s-t}
\frac{\beta^n}{n!} x^{1-p} = \int_0^{\infty} ds dt \, e^{-s-t}
\frac{2^{-n+p-1} (-s-t+1)^n (t-s)^{1-p}}{n!}.
\end{align}
Performing the change of variables $s \rightarrow  u v$, $t
\rightarrow  u - u v$, $\int_{0}^{\infty} ds dt \rightarrow
\int_{0}^{\infty} du \int_{0}^{1} dv \, u  $ into the above
integral (\ref{realplusbrstl07}), we obtain
\begin{align}
\int_0^{\infty} ds dt \, e^{-s-t} \frac{\beta^n}{n!} x^{1-p} &=
\int_{0}^{\infty} du \int_{0}^{1} dv \frac{e^{-u} 2^{-n+p-1}
(1-u)^n u^{2-p} (1-2 v)^{1-p}}{n!}
\nonumber \\
&= \frac{ \left((-1)^p-1\right) 2^{-n+p-2}}{(p-2) n!}
\int_{0}^{\infty} du \, e^{-u} (1-u)^n u^{2-p} \nonumber \\
\label{realplusbrstl08} &= \frac{ \left((-1)^p-1\right)
2^{-n+p-2}}{(p-2) n!}
 \mathcal{F}(n,2-p),
\end{align}
where we have defined
\begin{align}
\label{realplusbrstl09}
 \mathcal{F}(M,N) = \int_{0}^{\infty} du \, e^{-u} (1-u)^M u^{N}
 =\sum_{k=0}^{M} (-1)^{M-k} \binom{M}{k} (M+N-k)!
\end{align}

Proceeding in the same way, we can also calculate the curly
$\mathcal{L}_0$ level expansion of the first terms appearing on
the right hand side of equation (\ref{realplusbrstl02}). Adding up
all the results, we show that the string field
(\ref{realplusbrstlo}) has the following curly $\mathcal{L}_0$
level expansion
\begin{align}
\label{realclolevel1} \hat \Phi  = \sum_{n=0}^{\infty}
\sum_{p=-\infty}^{1} f_{n,p} \hat{\mathcal{L}}^n \tilde c_p
|0\rangle + \sum_{n=0}^{\infty} \sum_{p=-\infty}^{1}
\sum_{q=-\infty}^{1} f_{n,p,q} \hat{\mathcal{B}}
\hat{\mathcal{L}}^n \tilde c_p \tilde c_q |0\rangle,
\end{align}
where the coefficients $f_{n,p}$ and $f_{n,p,q}$ are given by
\begin{align}
\label{fnp} f_{n,p} &= \frac{\left(1-(-1)^p\right) 2^{-n+p-4}
\left(3
   \mathcal{F}(n,1-p)+\frac{1}{2-p}\mathcal{F}(n,2-p)\right)}{n!}, \\
\label{fnpq} f_{n,p,q} &= \frac{\left((-1)^q-(-1)^p\right)
2^{-n+p+q-6} \mathcal{F}(n,2-p-q)}{n!} .
\end{align}

To compute the kinetic term, we start by replacing the string
field $\hat\Phi$ with $z^{\mathcal{L}_0}\hat\Phi$, so that states
in the curly $\mathcal{L}_0$ level expansion will acquire
different integer powers of $z$ at different levels. As we are
going to see, the parameter $z$ is needed because we need to
express the kinetic term as a formal power series expansion if we
want to use Pad\'{e} approximants. After doing our calculations,
we will simply set $z=1$.

Let us start with the evaluation of the kinetic term as a formal
power series expansion in $z$. By inserting the expansion
(\ref{realclolevel1}) of the string field $\hat\Phi$ into the
kinetic term, and using the conservation laws studied in section 3
to evaluate the corresponding two point vertices, we obtain
\begin{align}
 \text{tr}[z^{\mathcal{L}_0} \hat\Phi \,
Q_B\big( z^{\mathcal{L}_0} \hat\Phi\big)] =&-\frac{4}{\pi ^2
z^2}+\big(1-\frac{2}{\pi ^2}\big)-z+\big(\frac{3}{2}-\frac{3 \pi
^2}{8}\big) z^2 +\big(-\frac{7}{2}+\frac{19 \pi
^2}{8}\big) z^3 \nonumber \\
&+\big(\frac{41}{4}-\frac{51 \pi ^2}{4}+\frac{\pi ^4}{8}\big)
z^4+\big(-36+\frac{279 \pi ^2}{4}-\frac{35 \pi ^4}{16}\big) z^5
\nonumber \\
\label{realclolevel2} &+\big(\frac{293}{2}-\frac{1615 \pi
^2}{4}+\frac{825 \pi ^4}{32}-\frac{5 \pi
   ^6}{128}\big) z^6+ \cdots
\end{align}

Considering terms up to order $z^6$, and setting $z = 1$, from
equation (\ref{realclolevel2}) we get $3328\%$ of the expected
result (\ref{Kin1}). In principle, we can compute the curly
$\mathcal{L}_0$ level expansion of the kinetic term up to any
desired order, however as we increase the order, the involved
tasks demand a lot of computing time. We have determined the
series (\ref{realclolevel2}) up to order $z^{18}$, and setting
$z=1$, we obtain about $1.5036\times 10^{15}\%$ of the expected
result. As we can see, if we naively set $z=1$ and sum the series,
we are left with a non-convergent result.

Recall that in numerical curly $\mathcal{L}_0$ level truncation
computations, a regularization technique based on Pad\'{e}
approximants provides desired results for gauge invariant
quantities like the energy
\cite{Erler:2009uj,Arroyo:2009ec,AldoArroyo:2011gx,Arroyo:2014pua}.
Let us see if after applying Pad\'{e} approximants, we can recover
the expected result.

To start with Pad\'{e} approximants, first let us define the
normalized value of the kinetic term as follows
\begin{align}
\label{realclolevel3}  \hat E(z) \equiv \frac{\pi ^2 z^2}{3}
\text{tr}[z^{\mathcal{L}_0} \hat\Phi \, Q_B\big( z^{\mathcal{L}_0}
\hat\Phi\big)].
\end{align}
Since the series for the kinetic term (\ref{realclolevel2}) is
known up to order $z^{18}$, we can write the series for $\hat
E(z)$ up to order $z^{20}$, and after considering a numerical
value for $\pi$, we obtain
\begin{align}
\hat E(z)=\sum E_k \, z^k =&-1.33333+2.6232 z^2-3.28987
z^3-7.24133 z^4+65.601 z^5-340.21 z^6 \nonumber \\&+ 1445.31
z^7-4489.28 z^8-1862.15 z^9+218120. z^{10}-2.84231\times 10^6
z^{11} \nonumber
\\&+ 2.83085\times 10^7 z^{12}-2.4607\times 10^8
z^{13}+1.87127\times 10^9 z^{14}\nonumber \\&-1.1131\times 10^{10}
z^{15}+ 1.91077\times 10^{10} z^{16}+9.10893\times 10^{11}
z^{17} \nonumber \\
\label{realclolevel4} &-2.20996\times 10^{13} z^{18}+
3.69796\times 10^{14} z^{19}-5.29538\times 10^{15} z^{20}.
\end{align}
In general, to construct a Pad\'{e} approximant of order
$P^n_{n}(z)$ for the normalized value of the kinetic term
(\ref{realclolevel3}), we need to truncate the series
(\ref{realclolevel4}) up to order $z^{2n}$.

As an illustration, let us compute the normalized value of the
kinetic term using a Pad\'{e} approximant of order $P^2_{2}(z)$.
First, we express $\hat E(z)$ as the rational function
$P^2_{2}(z)$
\begin{eqnarray}
\label{realclolevel5} \hat E(z)=P^2_{2}(z)=\frac{a_0+a_1z+a_2z^2
}{1+b_1z+b_2z^2}  \, .
\end{eqnarray}
Expanding the right hand side of (\ref{realclolevel5}) around
$z=0$ up to order $z^{4}$ and equating the coefficients of
$z^{0}$, $z^{1}$, $z^{2}$, $z^{3}$, $z^{4}$ with the expansion
(\ref{realclolevel4}), we get a system of algebraic equations for
the unknown coefficients $a_0$, $a_1$, $a_2$, $b_1$, and $b_2$.
Solving those equations we get
\begin{eqnarray}
a_0 = -1.3333, \; a_1=-1.6721, \; a_2=-3.1546 , \; b_1=1.2541, \;
b_2=4.3333.
\end{eqnarray}
Replacing the value of these coefficients inside the definition of
$P^2_{2}(z)$ (\ref{realclolevel5}), and evaluating this at $z=1$,
we get the following value
\begin{eqnarray}
\label{realclolevel6} P^2_{2}(z=1) = -0.935125008 .
\end{eqnarray}

The results of our calculations are summarized in table
\ref{realkineticresult1}. As we can see, the value of $\hat E(z)$
at $z=1$ by means of Pad\'{e} approximants confirms the expected
analytical result $\hat E(1) = \frac{\pi ^2}{3} \text{tr}[
\hat\Phi \, Q_B \hat\Phi] \rightarrow -1$. Although the
convergence to the expected answer gets irregular at $n=4$, by
considering higher level contributions, we will eventually reach
to the right value.

Using an alternative resummation technique, we would like to
confirm the expected answer for the normalized value of the
kinetic term. We have used a second method which is based on a
combination of Pad\'{e} and Borel resummation. We replace the
Borel transform of $\hat E(z)$, which is defined as $\hat
E(z)_{\text{Borel}} = \sum E_k z^k/k! $, by its Pad\'{e}
approximant $P_{n}^{n}(z)_{\text{Borel}}$ and then evaluate the
integral
\begin{eqnarray}
\label{padeborel} \widetilde P_{n}^{n}(z) = \int_{0}^{\infty} dt
\, e^{-t} \, P_{n}^{n}(z t)_{\text{Borel}}
\end{eqnarray}
at $z=1$. In the third column of table \ref{realkineticresult1},
we list the results obtained for $\hat E(1)$ by means of
Pad\'{e}-Borel approximations. Note that starting at the value of
$n=4$, Pad\'{e}-Borel does a little better than Pad\'{e}.

\begin{table}[ht]
\caption{The Pad\'{e} and Pad\'{e}-Borel approximation for the
normalized value of the kinetic term $\hat E(z)=\frac{\pi ^2
z^2}{3} \text{tr}[z^{\mathcal{L}_0} \hat\Phi \, Q_B\big(
z^{\mathcal{L}_0} \hat\Phi\big)]$ evaluated at $z=1$. The second
column shows the $P_{n}^{n}$ Pad\'{e} approximation. The third
column shows the corresponding $\widetilde P_{n}^{n}$
Pad\'{e}-Borel approximation. In the last column, $P^{2n}_0$
represents a trivial approximation, a naively summed series.}
\centering
\begin{tabular}{|c|c|c|c|c|}
\hline
  & $P^{n}_{n}$ & $\widetilde P^{n}_{n}$ & $P^{2n}_0$   \\
    \hline $n=0$ &  $-1.3333333333$ & $-1.3333333333$ & $-1.3333333333$ \\
\hline  $n=2$&  $-0.9351250080$  & $-0.6792579899$ & $-9.2413341787$ \\
\hline $n=4$ &  $-0.7462344772$  & $-0.9160629680$ & $-3327.8214730$  \\
\hline $n=6$ &  $-0.9803952323$  & $-0.9938587065$ & $2.56791\times 10^7$ \\
\hline $n=8$ &  $-0.9800827399$ & $-1.0020031889$ & $9.62763\times 10^9$ \\
\hline  $n=10$&  $-0.9997340118$ & $-1.0017620332$ & $-4.94676\times 10^{15}$  \\
\hline
\end{tabular}
\label{realkineticresult1}
\end{table}

\section{$L_0$ level expansion analysis of the real solution}
To expand the string field (\ref{realplusbrstl02}) in the Virasoro
basis of $L_0$ eigenstates, we are going to use the following
formulas
\begin{eqnarray}
\label{expanL01} e^{-t_1 K} c e^{-t_2K} B c e^{-t_3K} = \frac{r
\cos ^2\left(\frac{\pi x}{r}\right) \left(\pi  (r-2 y)-r \sin
\left(\frac{2 \pi y}{r}\right)\right)}{4 \pi^2 }\widetilde{U}_{r}
c\left(\frac{2 \tan \left(\frac{\pi
   x}{r}\right)}{r}\right)|0\rangle \;\;\; \nonumber \\
   +  \frac{r \cos ^2\left(\frac{\pi  y}{r}\right) \left(\pi  (r+2 x)+r \sin \left(\frac{2 \pi  x}{r}\right)\right)}{4 \pi^2 }
   \widetilde{U}_{r} c\left(\frac{2 \tan
   \left(\frac{\pi
   y}{r}\right)}{r}\right)|0\rangle \;\;\; \nonumber \\
+ \sum_{k=1}^{\infty}\frac{(-1)^{k+1} 2^{2 k-1}
\left(\frac{1}{r}\right)^{2 k-3} \cos ^2\left(\frac{\pi
x}{r}\right) \cos ^2\left(\frac{\pi  y}{r}\right)}{\left(4
k^2-1\right)
   \pi^2 } \widetilde{U}_{r}
   b_{-2k} c\left(\frac{2}{r} \tan \left(\frac{\pi
   x}{r}\right)\right) c\left(\frac{2}{r} \tan \left(\frac{\pi
   y}{r}\right)\right)|0\rangle , \; \\ r=t_1+t_2+t_3+1 , \;\;\;\;\;
x=\frac{1}{2}(t_3-t_1-t_2), \;\;\;\;\; y=\frac{1}{2}(t_2+t_3-t_1),
\;\;\;\;\;\;\;\;
\end{eqnarray}
where the operator $\widetilde{U}_{r}$ is defined as
\begin{eqnarray}
\label{expanL02} \widetilde{U}_{r} \equiv  \cdots e^{u_{10,r}
L_{-10}} e^{u_{8,r} L_{-8}} e^{u_{6,r} L_{-6}} e^{u_{4,r}
L_{-4}}e^{u_{2,r} L_{-2}}.
\end{eqnarray}
To find the coefficients $u_{n,r}$ appearing in the exponentials,
we use
\begin{align}
\frac{r}{2} \tan (\frac{2}{r} \arctan z) &= \lim_{N \rightarrow
\infty} \big[f_{2,u_{2,r}} \circ  f_{4,u_{4,r}} \circ
f_{6,u_{6,r}} \circ f_{8,u_{8,r}} \circ
f_{10,u_{10,r}} \circ \cdots \circ f_{N,u_{N,r}}(z)\big] \nonumber \\
&= \lim_{N \rightarrow \infty}\big[ f_{2,u_{2,r}} ( f_{4,u_{4,r}}
( f_{6,u_{6,r}} ( f_{8,u_{8,r}} ( f_{10,u_{10,r}}(\cdots
(f_{N,u_{N,r}}(z)) \dots )))))  \big],
\end{align}
where the function $f_{n,u_{n,r}}(z)$ is given by
\begin{eqnarray}
f_{n,u_{n,r}}(z) = \frac{z}{(1-u_{n,r} n z^n)^{1/n}}.
\end{eqnarray}

Employing the set of equations (\ref{expanL01})$-$(\ref{expanL02})
for the string field (\ref{realplusbrstl02}), we obtain
\begin{eqnarray}
\label{expanL03} \hat \Phi = \int_{0}^{\infty} dt \frac{e^{-t} r
\sin ^2\left(\frac{\pi }{2 r}\right) \left(2 \pi  r-r\sin
\left(\frac{\pi }{r}\right)+\pi
   \right)}{16 \pi ^2} \widetilde{U}_{r} \Big(c\big(-\frac{2 \tan \left(\frac{\pi  t}{2 r}\right)}{r}\big)+c\big(\frac{2 \tan \left(\frac{\pi  t}{2
   r}\right)}{r}\big)\Big) \nonumber \\
 +  \int_{0}^{\infty} dt \sum_{k=1}^{\infty} \frac{e^{-t}(-1)^{k+1} 2^{2 k-3} \left(\frac{1}{r}\right)^{2 k-3} \sin ^4\left(\frac{\pi }{2 r}\right)}{\pi ^2 \left(4
   k^2-1\right)} \widetilde{U}_{r}
 b_{-2k} c\big(-\frac{2 \tan \left(\frac{\pi  t}{2 r}\right)}{r}\big)c\big(\frac{2 \tan \left(\frac{\pi  t}{2
   r}\right)}{r}\big) \nonumber \\
   + \int_0^{\infty} ds \int_0^{\infty} dt  \frac{e^{-s-t}(1+s+t)^2 \cos ^2\left(\frac{\pi  (t-s)}{2 (1+s+t)}\right)}{8 \pi } \widetilde{U}_{1 + s + t} c\left(\frac{2 \tan \left(\frac{\pi  (t-s)}{2
   (1+s+t)}\right)}{1+s+t}\right), \;\;\;\;\;\;\;\;
\end{eqnarray}
where $r=1+t$.

By writing the $c$ ghost in terms of its modes $c(z)=\sum_{m}
c_m/z^{m-1}$ and employing equations (\ref{expanL02}) and
(\ref{expanL03}), the string field $\hat \Phi$ can be readily
expanded and the individual coefficients can be numerically
integrated. For instance, let us write the expansion of $\hat
\Phi$ up to level fourth states
\begin{align}
\label{expanL04} \hat \Phi  = & 0.45457753 c_1 |0\rangle +
0.17214438 c_{-1} |0\rangle -0.03070678 L_{-2} c_{-1} |0\rangle
-0.01400692 b_{-2} c_0 c_1 |0\rangle \nonumber \\ & -0.00605891
L_{-4} c_1|0\rangle + 0.02033379 L_{-2} L_{-2} c_1|0\rangle +
0.16194599 c_{-3}|0\rangle  \nonumber \\ & -0.00976204 b_{-2}
c_{-2} c_{1}|0\rangle -0.01053192 L_{-2} c_{-1}|0\rangle +
 0.00976204 b_{-2} c_{-1}c_{0}|0\rangle \nonumber
\\ &  + 0.00465417 b_{-4} c_{0} c_{1}|0\rangle -0.00308797 L_{-2} b_{-2} c_{0}
c_{1}|0\rangle + \cdots.
\end{align}

As in the case of the curly $\mathcal{L}_0$ level expansion
analysis, to evaluate the normalized value of the vacuum energy,
first we perform the replacement $\hat \Phi \rightarrow z^{L_0}
\hat \Phi$ and then using the resulting string field $z^{L_0} \hat
\Phi$, we define, the analogue of equation (\ref{realclolevel3})
\begin{align}
\label{expanL05}  \tilde E(z) = \frac{\pi ^2 z^2}{3}
\text{tr}[z^{L_0} \hat\Phi \, Q_B\big( z^{L_0} \hat\Phi\big)].
\end{align}
The normalized value of the vacuum energy is obtained just by
setting $z = 1$. Since the kinetic term is diagonal in $L_0$
eigenstates, the coefficients of the energy (\ref{expanL05}) at
order $z^{2L}$ are exactly the contributions from fields at level
$L$. We have expanded the string field $\hat \Phi$ given in
equation (\ref{expanL03}) up to level twelfth states, and hence
the series of $\tilde E(z)$ can be determined up to the order
$z^{24}$
\begin{align}
\label{expanL06}  \tilde E(z) = &
-0.6798207-0.1505669z^4-0.2304622z^8+0.238568z^{12}-0.3465834z^{16}
\nonumber \\ &+0.4456892 z^{20}-0.58817204 z^{24}.
\end{align}

If we naively evaluate the truncated vacuum energy
(\ref{expanL06}), i.e., setting $z = 1$ in the series before using
Pad\'{e} or Pad\'{e}-Borel approximations, we obtain a
non-convergent result. Note that the series (\ref{expanL06}) is
less divergent than the series (\ref{realclolevel4}) that has been
obtained in the case of the curly $\mathcal{L}_0$ level expansion
analysis of the energy.

Let us re-sum the divergent series (\ref{expanL06}). To obtain the
Pad\'{e} or Pad\'{e}-Borel approximation of order $P^n_n$ for the
energy, we will need to know the series expansion of $\tilde E(z)$
up to the order $z^{2n}$. The results of these numerical
calculations are summarized in table \ref{realkineticresult1L0}.

\begin{table}[ht]
\caption{The Pad\'{e} and Pad\'{e}-Borel approximation for the
normalized value of the vacuum energy $\tilde E(z)=\frac{\pi ^2
z^2}{3} \text{tr}[z^{L_0} \hat\Phi \, Q_B\big( z^{L_0}
\hat\Phi\big)]$ evaluated at $z=1$. The second column shows the
$P_{n}^{n}$ Pad\'{e} approximation. The third column shows the
corresponding $\widetilde P_{n}^{n}$ Pad\'{e}-Borel approximation.
In the last column, $P^{2n}_0$ represents a trivial approximation,
a naively summed series.} \centering
\begin{tabular}{|c|c|c|c|c|}
\hline
  & $P^{n}_{n}$ & $\widetilde P^{n}_{n}$ & $P^{2n}_0$   \\
    \hline $n=0$ &  $-0.6798207586$ & $-0.6798207586$ & $-0.6798207586$ \\
\hline  $n=4$&  $-0.3960692519$  & $-0.8200429863$ & $-1.0608500359$ \\
\hline $n=8$ &  $-0.9687853277$  & $-0.9606366393$ & $-1.1688654612$  \\
\hline $n=12$ &  $-0.9782343686$  & $-0.9537697224$ & $-1.3113482982$ \\
\hline
\end{tabular}
\label{realkineticresult1L0}
\end{table}

\section{Summary and discussion}
We have analyzed the validity of the recently proposed real
tachyon vacuum solution \cite{Jokel:2017vlt}, in open bosonic
string field theory. We have found that the solution solves in a
non trivial way the equation of motion when contracted with
itself. Let us point out that a similar test of consistency was
performed by Okawa \cite{Okawa:2006vm}, Fuchs, Kroyter
\cite{Fuchs:2006hw} and Arroyo \cite{Arroyo:2009ec} for the case
of the original Schnabl's solution \cite{Schnabl:2005gv}.

As a second test of consistency, we have analyzed the solution
from a numerical point of view. Using either the curly
$\mathcal{L}_0$, or the Virasoro $L_0$ level expansion of the
solution, we have found that the expression representing the
energy is given in terms of a divergent series, which nevertheless
can be re-summed, either by means of Pad\'{e} technique or a
combination of Pad\'{e}-Borel resummation to bring the expected
result in agreement with Sen's conjecture.

It would be interesting to analyze other real solutions. For
instance, the tachyon vacuum solution corresponding to the
regularized identity based solution \cite{Zeze:2010sr}. The real
version of this solution, obtained by means of a similarity
transformation, contains square roots and consequently the
analytical and numerical computations of the energy become
cumbersome \cite{Arroyo:2010sy,AldoArroyo:2011gx}. Employing the
prescription studied in reference \cite{Jokel:2017vlt}, it should
be possible to find an alternative real version for this
regularized identity based solution.

Finally, regarding to the modified cubic superstring field theory
\cite{Arefeva:1989cp} and Berkovits non-polynomial open
superstring field theory \cite{Berkovits:1995ab}, since these
theories are based on Witten's associative star product, their
mathematical setup shares the same algebraic structure of the open
bosonic string field theory, and thus the prescription developed
in reference \cite{Jokel:2017vlt} and the results shown in this
paper should be extended to construct and study new real solutions
in the superstring context like the ones discussed in references
\cite{Arroyo:2014pua,Arroyo:2013pha,AldoArroyo:2012if,Arroyo:2016ajg,Erler:2007xt,Erler:2010pr,Erler:2013wda}.

\section*{Acknowledgements}
I would like to thank Ted Erler and Max Jokel for useful
discussions.


\end{document}